\documentclass{iopart}

\usepackage{graphicx}
\usepackage[utf8]{inputenc}
\usepackage[T1]{fontenc}

\usepackage{color}
\usepackage{iopams}
\begin{document}

\title[Expansion of a spherical plasma cloud]
{On the unconstrained expansion of a spherical plasma cloud turning
collisionless : case of a cloud generated by a nanometer dust grain 
impact on an uncharged target in space.}

\author{F Pantellini$^1$, S Landi$^2$, A Zaslavsky$^1$ and N Meyer-Vernet$^1$}

\address{1 LESIA, Observatoire de Paris, CNRS, UPMC,
Universit\'e Paris Diderot;\\ 5 Place Jules Janssen, 92195
Meudon, France}
\address{2 Dipartimento di Astronomia e
Scienza dello Spazio, Largo Enrico Fermi 2, 50125 Firenze, Italy}

\eads{\mailto{Filippo.Pantellini@obspm.fr}}

{Published in \underline{\PPCF}}

\begin{abstract}
Nano and micro meter sized dust particles travelling
through the heliosphere at several hundreds of km/s
have been repeatedly detected by interplanetary
spacecraft.
When such fast moving dust particles hit a solid target 
in space, an expanding plasma cloud is formed 
through the vaporisation and ionisation of the dust particles
itself and part of the target material at and near the
impact point. Immediately after the impact the small and dense
cloud is dominated by collisions and the expansion
can be described by fluid equations. However,
once the cloud has reached $\mu{\rm m}$ dimensions, the
plasma may turn collisionless and
a kinetic description is required to describe
the subsequent expansion.
In this paper we explore the late and possibly collisionless 
spherically symmetric unconstrained expansion of a single ionized
ion-electron plasma using N-body simulations.
Given the strong uncertainties concerning the
early hydrodynamic 
expansion, we assume that at the time of the transition to the collisionless
regime the cloud density and temperature are spatially uniform. 
We do also neglect the role of the ambient plasma.  
This is a reasonable assumption
as long as the cloud density is substantially higher than the
ambient plasma density.
In the case of clouds generated by fast 
interplanetary dust grains hitting a solid target 
some $10^7$ electrons and ions are liberated 
and the in vacuum approximation is
acceptable up to meter order cloud dimensions.
As such a cloud can be estimated to become collisionless
when its radius has reached $\mu$m order dimensions,
both the collisionless approximation and the in vacuum
approximation are expected to hold during a long lasting phase 
as the cloud grows by a factor $10^6$.
With these assumptions, we find that the transition from the collisional to
the collisionless regime could occur
when the electron Debye length $\lambda_{\rm D}$ within the cloud 
is much smaller than the cloud radius $R_0$, i.e. 
$\Lambda\equiv\lambda_{\rm D}/R_0\ll 1$.
This implies a quasi-neutral expansion regime where the radial 
electron and ion density profiles are equal through most of the cloud
except at the cloud-vacuum interface. 
The consequence of $\Lambda$ being much smaller that unity
implies that the electrostatic fields within a cloud generated
by a dust impact on a neutral target is $\sim 100$ times weaker than 
in the case of grains hitting a spacecraft, where the positive 
potential of the target is strong enough to strip-off all the electrons 
from the expanding cloud leading to a "Coulomb explosion" like regime 
(e.g. Peano et al (2007) \cite{Peano_Martins_al_2007}).

\end{abstract}

\maketitle

\section{Introduction\label{sec_intro}}

The problem of the expansion of a plasma into vacuum has
received much attention in recent years mainly
in the context of the understanding of the expansion
of plasma clouds generated by laser irradiated 
materials
\cite{Ditmire_Zweiback_al_1999,Ter-Avetisyan_Schnurer_al_2001,
Kishimoto_Masaki_al_2002,Mora_2005, Breizman_Arefiev_al_2005}.
The expansion of negatively charged dust particles 
in cometary tails \cite{Lonngren_1990,Pillay_Singh_al_1997} 
and the expansion of the solar wind plasma
into the wake region of inert objects such as asteroids or the
moon \cite{Birch_Chapman_2002} has also stimulated theoretical and numerical
studies on the problem of the expansion of a plasma into vacuum.
The impact of fast moving clusters of atoms or molecules
on a solid surface are also known to produce expanding plasma
clouds. In particular, dust particles, typically in the micro to nano meter
range, hitting spacecraft at velocities up to hundreds of km/s
have been repeatedly detected in space
\cite{Gurnett_Grun_al_1983,Aubier_Meyer-Vernet_al_1983,Utterback_Kissel_1990,
Zook_Grun_al_1996,
McBride_McDonnell_1999, Mann_Czechowski_2005,Kempf_Srama_al_2005,
Meyer-Vernet_Maksimovic_al_2009, Cyr_Kaiser_al_2009}.
In most laser plasmas experiments only the electrons (not the ions) are
heated by the laser's electromagnetic field. In these experiments,
the initial state of the plasma is characterised by
of hot electron population, carrying all of the energy, 
and a cold ion population, too
tenuous for collisions to operate. On the other hand,
dust impact generated plasma clouds are initially dense enough  
for electrons and ions to be in thermodynamic equilibrium 
everywhere within the cloud with the possible
exception of the cloud/vacuum interface.
One fluid models
\cite{Zeldovich_Raizer_2002,Pack_1953,Stanyukovich_1960,
Tzuk_Barmashenko_al_1993},
or two fluid models allowing for a separate description of ions and electrons 
(see the classical review paper \cite{Sack_Schamel_1987})
are the appropriate tools to model the collisional regime 
of the expansion.
At some point however, provided the expansion takes place in
vacuum or in a tenuous plasma and provided the collisional mean free
path of an electron grows faster than the cloud radius $R$,
the expansion becomes collisionless
and a kinetic description necessary. Typically
a nano meter dust particle impact at the solar wind speed is expected to
ignite an expanding plasma cloud with some $10^7$ electrons and
ions and a characteristic temperature of 10eV  
turns collisionless at cloud dimensions $R \gtrsim \mu{\rm m}$. 
The cloud then continues its expansion in the free collisionless regime
until its density has declined to a value comparable
to the ambient plasma density, which at Earth's orbit occurs
for a meter order cloud radius. 
Thus, during the free collisionless expansion regime the cloud radius
grows by a factor of order $10^6$  before it merges 
with the ambient plasma. At such small scales 
magnetic forces can be safely neglected given 
that the Larmor radius of a low energy 1eV electron in interplanetary   
space already exceeds 600 m.  

The aim of this paper is to explore the unconstrained collisionless
expansion of a plasma cloud using kinetic simulations. 
The plasma made of identical single ionised ions and an equal number of
electrons is assumed to be initially confined within a spherical shell 
which will be instantly removed to let the plasma expand freely into vacuum. The
plasma is assumed to be initially at rest and at thermodynamic equilibrium
implying equal ion and electron temperatures. 
This kind of initial condition differs from most of the published 
papers on the expansion of a plasma into vacuum where ions
are generally assumed to be cold as in most laser heated 
laboratory plasma (e.g.   
\cite{Mora_2003, Murakami_Basko_2006, Kumar_Pukhov_2008,
Beck_Pantellini_2009} using fluid models, or \cite{Peano_Peinetti_al_2006,
Peano_Martins_al_2007} using kinetic models).
We note that some of the presented  results,  
in particular concerning the shape of the asymptotic electron velocity distribution 
function (see Figure \ref{fig5}), have been anticipated 
by Manfredi and Mola (1993) \cite{Manfredi_Mola_al_1993} 
using the strictly collisionless Vlasov model and a hotter plasma. 
In this work we use a one dimensional N-body scheme to solve the equations of motion 
for a large number of ions and electrons so that collisions are not exclude a priori. 
The similarities between our results and the results by Manfredi and Mola 
are then due to the fact that we assume the cloud to be marginally collisional 
from the beginning with the expansion further reducing the collisionality 
until an asymptotic ``frozen'' self-similar state is reached. 
It is worth noting that analytic self-similar solutions in the
quasi-neutral limit, where charge separation is assumed to be small exist in the literature 
(see e.g. \cite{Kumar_Pukhov_2008, Dorozhkina_Semenov_1998}). Such solutions are expected to hold in
the limit of an electron Debye length being small with respect to the cloud dimension which we do
suggest to be the case if the latter turns collisionless during its expansion. However, the
formation of an extended electron precursor, which appears to be inevitable in the spherical case 
\cite{Murakami_Basko_2006, Beck_Pantellini_2009}, suggest that the quasi-neutral assumption
necessarily fails in the electron dominated, outer shells of the cloud.

The plan of the paper is as follows. 
In Section \ref{sec_def} we introduce the basic parameters and equations
relevant to the problem of the expansion of a 
spherically symmetric plasma. Since we focus our attention to the case 
of clouds which are initially collision dominated, we complete 
Section \ref{sec_def} with an outline the
necessary conditions for a transition towards a collisionless regime to occur.
In section \ref{sec_setup} the initial conditions
and parameters for a simulation of a typical case are specified.
The results of the simulation are discussed in detail in
sections \ref{sec_asymptotic} to \ref{sec_velocities}.
A summary of the paper is presented in section \ref{sec_conclusions}.

\section{Definition of the problem\label{sec_def}}
We consider a spherically symmetric electron-ion
plasma cloud made of $N/2$ singly charged ions and $N/2$ electrons
expanding into vacuum.
During the first phase
of the expansion the plasma is supposed to be sufficiently dense
to be dominated by interparticle collisions.
This phase is conveniently described by
fluid dynamics
\cite{Zeldovich_Raizer_2002,Stanyukovich_1960,Tzuk_Barmashenko_al_1993,
Sack_Schamel_1987} and will not be
treated in this paper. Under favourable conditions, however,
the cloud radius $R(t)$ may grow larger than
the collisional mean free path of a thermal electron.
At this particular radius $R=R_0$ 
the cloud plasma becomes collisionless and enters a new
regime which is no longer tractable within the frame
of a fluid theory. The reason for using electrons instead of 
ions to define the end of the collisional regime is that for a given 
temperature the collisionality of the latter may be substantially 
reduced as soon as the external shells of the cloud start moving faster than 
the ion thermal velocity as under such circumstances ions can no longer approach each other. 
As we shall see below the expansion velocity is indeed suprathermal for the ions 
after a short lapse of time roughly corresponding to the time required 
for the cloud to double its radius.

\subsection{Initial state of the cloud\label{sub_initial_state}}

As the expansion is supposed to be collisional 
for $R<R_0$ we assume that electrons and ions
are initially (at time $t_0$) in the state of
thermodynamic equilibrium and confined within the spherical shell 
$R=R_0$.
As the state of the plasma at the end of the
fluid (collision dominated) phase is generally not
known as it strongly depends on the cloud structure 
at the time of its formation and also on the
adopted equation of state, we assume that ions and electrons are
initially distributed according to a zero mean velocity  
Maxwell-Boltzmann distribution with temperature $T_0$.
In order for the electron mean free path to be
uniquely defined we do also assume that at $t=0$ 
the density within the cloud is spatially uniform.
A more realistic description of the initial state of the
cloud when $R=R_0$, not even
taking into account the presence of different ion species
and neutrals (e.g. \cite{Hornung_Kissel_1994}), should include a
non zero, radially varying fluid velocity profile, and a complex
plasma-vacuum interface (see \cite{Sack_Schamel_1987}).
We note that even if one assumes that the early,
collisional phase is governed by simple inviscid
gas dynamics, the spatial and temporal structure of the expanding cloud
has been shown to strongly depend on both the assumed energy equation
(isentropic, isothermal, ...) and on the  density
and temperature distribution within the cloud at the time of its 
formation \cite{Stanyukovich_1960, Zeldovich_Raizer_2002,
Tzuk_Barmashenko_al_1993}. 
A few words on the radial velocity profile $u(r,t)$.
In situations where the cloud radius $R$ is allowed
to grow much larger than the radius of the cloud at the time of its formation
the evolution must be close to self-similar.
Unlike the density and temperature profiles, 
the velocity profile then converges towards 
$u=r/t$ for $t\rightarrow \infty$ independently of the 
conditions at the time of formation and independently of whether
the governing equations are fluid \cite{Stanyukovich_1960,
Zeldovich_Raizer_2002, Tzuk_Barmashenko_al_1993, Molmud_1960} or kinetic
\cite{Peano_Martins_al_2007,Murakami_Basko_2006,Beck_Pantellini_2009}.
One may then be tempted to select a linear velocity
profile $u(r,t=0)=r/\delta t$ as initial condition for the collisionless
regime, where $-\delta t$ is the instant of cloud formation. Unfortunately, such a
profile is a priori incompatible
with the assumption of a constant density profile
unless very special, and unlikely, conditions exist at $t=-\delta t$.
More sophisticated initial conditions based on
approximate self-similar solutions from compressible
gas dynamics (e.g. \cite{Zeldovich_Raizer_2002,
Tzuk_Barmashenko_al_1993}) will be discussed in a future publication.

\subsection{Basic parameters \label{sub_parameters}}

Previous works
\cite{Murakami_Basko_2006,Peano_Martins_al_2007,Beck_Pantellini_2009}
on the spherical expansion of a plasma into vacuum
have pointed out that the problem is characterized by
the dimensionless parameter $\Lambda\equiv\lambda_{\rm D}/R_0$ at $t=0$,
where  $\lambda_{\rm D}$ is the electron Debye length (SI units)
\begin{equation}
 \lambda_{\rm D} = \left(\frac{\varepsilon_0 T}{n e^2
}\right)^{1/2}\label{eq_ld}. 
\end{equation}
In equation (\ref{eq_ld})   
$e$ is the absolute value of the electron charge, $\varepsilon_0$ the 
permittivity of free space, $T$ the
temperature and $n$ the electron density. In equation (\ref{eq_ld})
and during the remaining of this paper we assume that temperatures
are given in energy units, i.e. temperatures are implicitly multiplied  
by the Boltzmann constant $k_{\rm B}$.
In situations when $\Lambda\ll 1$ the thermal energy of the electrons is too
low
to allow for a substantial charge separation at the cloud surface:
the expansion is quasi-neutral. On the other hand, in situations when  
$\Lambda\gg 1$ most 
of the electrons are energetic enough to overcome the
electrostatic forces which bind them to the ions.
In this case, the cloud becomes positively charged on a time scale of
the order of $R_0/v_{\rm e}$ ($v_{\rm e}\equiv (2T/m_{\rm e})^{1/2}$  is
the electron thermal velocity) and the associated peak electric field
is much stronger than in the quasi-neutral case.
In the limit $\Lambda\rightarrow \infty$ (the so called Coulomb explosion)
all electrons escape from the cloud and the expansion is driven mainly by the
repulsive forces pushing the unshielded ions away from each other. 

Let us now estimate the parameter $\Lambda$ at time $t=0$, 
when the plasma becomes collisionless. To this end we use
the Fokker-Planck expression for the mean free path
of a thermal electron
\begin{equation}
 l_{\rm e} = 16\pi\varepsilon_0^2 \:\frac{T^2}{e^4 n
\lambda}\label{eq_mfp}
\end{equation}
where $\lambda\equiv \ln (\lambda_{\rm D}/r_{\rm s})$ is the Coulomb logarithm
and $r_{\rm s}$ the strong interaction radius, usually defined as the larger of  
the classical distance for a strong electrostatic interaction between thermal electrons
$e^2/(12\pi\varepsilon_0 T)$ or the de Broglie length for a thermal electron 
$\hbar/(3 m_{\rm e}T)^{1/2}$, where $\hbar$ is the reduced Planck constant. 
For temperatures exceeding 9eV the quantum mechanical definition should therefore be used 
to define the $r_{\rm s}$. As typical cloud temperatures are expected to be of the order 
of a few eV up to at most 20eV and also because of the classical nature of the 
presented simulation, we stick to the classical definition throughout the paper. 
We emphasise that this assumption does not invalidate the subsequent discussions 
and the presented simulation for the case of temperatures higher than 9eV since  
we do only require $r_{\rm s}$ being small  
with respect to the radius of the spherical shell $r_{\rm min}$ 
defining the inner boundary of the simulation domain (see \ref{sec_setup}). 
Now, even for an exceedingly hot cloud at 81eV, the quantum mechanical 
definition of $r_{\rm s}$ is just three times larger than the classical definition.

Equation (\ref{eq_mfp}) is a good estimate of the
mean free path of a thermal electron in a weakly coupled plasma
with $\lambda\gtrsim 10$. For values $2 \lesssim \lambda \lesssim 10$
equation (\ref{eq_mfp}) may still be used as a fair estimate
of $l_{\rm e}$.
By noting that the initial density $n_0$ is related to the cloud radius
and the total number of electrons $N/2$ via
\begin{equation}
 n_0=\frac{N/2}{\displaystyle\frac{4\pi}{3} R_0^3}
\end{equation}
it follows from (\ref{eq_ld}) and by setting
$l_{\rm e}(t=0)=R_0$ in (\ref{eq_mfp}) that
the dimensionless parameter $\Lambda$ only depends on the
total number of charged particles $N$ in the cloud, viz. 
\begin{equation}
 \Lambda \equiv \frac{\lambda_{\rm D}}{R_0} =
\left[\frac{\lambda(N)}{6 N}\right]^{1/4} .\label{eq_lambda}
\end{equation}
We note indeed, that given the constraint $R_0=l_{\rm e}(t=0)$, 
the  Coulomb logarithm $\lambda$ is a function of the 
total number of particles $N$ via
$6N = (4/3)^4\:e^{4\lambda}\lambda^{-3}$ 
which leads to the relation $\Lambda=0.75\lambda e^{-\lambda}$.
Equation (\ref{eq_lambda}) indicates that the dimensionless parameter $\Lambda$
is independent of the temperature $T_0$ and 
much smaller than unity as $N$ is generally a large number
and $\lambda \lesssim 6$  for $N\lesssim 10^8$.
We conclude that at the time an initially collisional
plasma cloud becomes collisionless it finds itself
in the quasi-neutral expansion regime $\Lambda\ll 1$. For example,
in the case of a $10^{-20}\:{\rm kg}$ dust particle impacting on a spacecraft  
at solar wind speed the generated
plasma cloud contains some $N=10^7$ charged particles corresponding
to a Coulomb logarithm $\lambda \approx 5.5$ and $\Lambda=0.017$.

\subsection{From collisional to collisionless}
In this paper we restrict our discussion to the unconstrained
expansion of a plasma cloud where the
initial cloud's radius $R_0$ increases by a large
factor $R(t)/R_0 \gg 1$ before the dynamics of the expansion
becomes affected by external factors, such as the ambient plasma.
In the case of a dust impact generated plasma
expanding into the ambient plasma (the interplanetary plasma) 
we can neglect the ambient plasma as long as the cloud
density is much higher than the ambient plasma density. 
Indeed, for the nanodust impact considered 
above producing $N\approx 10^7$ charges with a typical
per particle
energy of 10eV (see \cite{Meyer-Vernet_Maksimovic_al_2009}) one finds, by
setting
$l_{\rm e}=R_0$ in equation (\ref{eq_mfp}), that the plasma cloud
can be considered collisionless for $R_0$ larger than a few $\mu{\rm m}$.
Since the initial size of the cloud, just after impact,
is necessarily comparable to the size of the dust grain
(at most a few tens nm) most of the plasma within the cloud 
is collisional at least during the first phase
of the expansion. The growing cloud will then turn collisionless
provided the mean free path $l_{\rm e}$ grows faster than the cloud radius
$R$. 
Disregarding the weak dependence of the Coulomb logarithm $\lambda$ on the
plasma density $n$ and assuming a polytropic equation of state
$T \propto n^{\gamma -1}$ one finds $l_{\rm e}/R \propto n^{2(\gamma -4/3)}$.
Thus, for $\gamma < 4/3$, 
the ratio $l_{\rm e}/R$ is a growing function of $R$ which means that the 
plasma must turn collisionless during expansion. If during the early
phase of the expansion the conductive heat flux $Q$ is unimportant the
expansion must be adiabatic with an index $\gamma_{\rm a}=5/3$ and the cloud
turns increasingly collisional, at least as long as standard adiabatic fluid
equations are applicable. Let us estimate under which conditions
the conductive heat flux is dominant by comparing the the conductive term and
the adiabatic term in the energy equation for a spherically symmetric 
collisional gas of point particles:
\begin{equation}
 \frac{3}{2}\frac{DT}{Dt} = 
-\frac{T}{r^2}\frac{\partial}{\partial r}(r^2 u) -
\frac{1}{n\:r^2}\frac{\partial}{\partial r}(r^2
Q)\label{eq_dTdt} 
\end{equation}
where $D/Dt\equiv \partial/\partial t + u\partial/\partial r$ is the 
material derivative. If the plasma in the cloud is collisional we can then use
the
Spitzer-Härm expression \cite{Spitzer_Harm_1953} for the conductive flux.
Neglecting the contribution to the flux from the ions we therefore set 
\begin{equation}
 Q=-1.6p_{\rm e} v_{\rm e}\:\frac{l_{\rm e}}{T} \frac{\partial T}{\partial
r}\label{eq_SH}
\end{equation}
where $p_{\rm e}=n_{\rm e}T$ is the electron pressure.
Let us concentrate on the centre of the cloud at $r=0$. If the plasma was
initially uniform, at rest and spherically symmetric, it follows that density,
pressure and temperature must have an extremum at $r=0$ and a vanishing first
derivative at all times. On the other hand if the fluid velocity was initially
zero at the centre it has to stay so for ever given that the acceleration is 
$\partial u/\partial r = -\varrho^{-1}\partial p/\partial r=0$ by virtue of the 
vanishing first derivative of the pressure.   
One may Taylor expand the velocity near $r=0$ as $u(r)=u'(0)r+\mathcal{O}(r^2)$,
where $^\prime\equiv \partial/\partial r$. According to the above
discussion, the Taylor expansion of the temperature up to the first non constant
term is $T(r)=T(0)+T''(0) r^2 /2+...$. Of course, the same expansion can be
applied to both density and pressure. To lowest order in $r$ we can then write
the energy equation (\ref{eq_dTdt}) for the central region of the cloud as  
\begin{equation}
 \frac{3}{2}\:\frac{\partial T_0}{\partial t} = 
      - 3T_0 u'(0) 
      + 1.6 v_{\rm e} l_{\rm e} T''(0) \label{eq_dTdt_0}
\end{equation}
where $T_0\equiv T(0)$. 
The first term on the right in equation (\ref{eq_dTdt_0}) corresponds to the
adiabatic cooling due to the expansion while the second term corresponds to the
non adiabatic cooling due to heat conduction. For the expansion to be
non adiabatic the latter has to be of comparable order, or larger, than the
former.
In this case the effective polytropic index of the plasma is smaller than
adiabatic $\gamma < 5/3$ and a transition from collisional to non collisional 
becomes possible if conduction is strong enough to reduce $\gamma$ below the
critical value $4/3$. In order to estimate the relative importance of the 2
terms in (\ref{eq_dTdt_0}) we need an estimate of $u'(0)$ and $T''(0)$. Using  
the available macroscopic parameters of the cloud, like its radius $R$ and the
expansion velocity $u_{\rm F}$ of the front,   it is natural to set
$u'(0)\approx u_{\rm F}/R$ and $T''(0)\approx T_0/R^2$. 
We can then estimate the departure from adiabaticity by comparing the
conductive to the adiabatic term, viz
\begin{equation}
  \frac{\rm non\;adiabatic\; term}{\rm adiabatic\; term } \approx
\frac{1.6}{3}\:\frac{v_{\rm e}}{u_{\rm F}}\: \frac{l_{\rm e}}{R}.
\label{eq_adiabatic_ratio}
\end{equation}
From (\ref{eq_adiabatic_ratio}) it appears that the expansion is
adiabatic in the limit of vanishing small mean free path $l_{\rm e}/R
\rightarrow 0$. However given that the typical expansion velocity   
$u_{\rm F}$ is of the order of a few times the ion thermal velocity, 
the ratio $v_{\rm e}/u_{\rm F}=\mathcal{O}([m_{\rm i}/m_{\rm e}]^{1/2})\gg 1$ is
much larger than unity. Thus, unless the Knudsen number $l_{\rm e}/R\ll
10^{-2}$ conduction is not negligible and the expansion cannot be adiabatic.   
Clouds with negligible heat conduction, i.e. for $l_{\rm e}/R\ll 10^{-2}$ much
smaller than $(m_{\rm e}/m_{\rm i})^{1/2}$, are expected both to remain
collisional and to obey the equations of standard adiabatic gas dynamics in
spherical geometry \cite{Landau_Lifshits_1987,Zeldovich_Raizer_2002}. 
This conclusion has to be softened somewhat 
as the collisional approximation for $Q$ used in
equation (\ref{eq_SH}) may be inaccurate 
for $l_{\rm e}/R\gtrsim 10^{-3}$ (e.g. \cite{Shoub_1983})  
and is even expected to saturate at $Q_{\rm sat} \approx 0.2 pv_{\rm
e}$ for $l_{\rm e}/R \gtrsim
0.1$ \cite{Luciani_Mora_al_1983,Salem_Hubert_al_2003}. 

\subsection{Problem reduction}

In the remaining of the paper we assume that the cloud under consideration 
goes through a phase where the Knudsen number $l_{\rm e}/R\gtrsim 10^{-2}$ 
ensuring that the cloud turns collisionless at some critical radius $R=R_0$.
The free expansion then continues 
until the cloud's density has decreased to a level comparable to
the ambient plasma.   
At Earth orbit, where the solar wind density is generally smaller than 
10 electrons per ${\rm cm}^{-3}$, the cloud density is substantially
larger than the ambient plasma density for up to a meter order
cloud radius $R$. Therefore, the expansion can be assumed
to be both collisionless and independent of the ambient plasma
while the cloud radius $R$ grows from $R_0 \sim \mu{\rm m}$ up to 
$R\sim {\rm m}$, which corresponds to an expansion factor
$R/R_0=\mathcal{O}(10^6)$.
Given such a large expansion factor it is justified to 
assume that all particles within the cloud
have purely radial velocities $\vec{v}=v\vec{r}/r$
as the transverse velocity component $v_{\perp}$ (perpendicular to the radial
direction) rapidly decreases during expansion since the angular
momentum of individual particles $L\equiv m r v_{\perp} $ is conserved in 
a collisionless and spherically symmetric field. 
Neglecting the centrifugal force due to the transverse component of the 
particle velocity,  the equations of motion for a particle of mass $m$ and
charge $q$
in a spherically symmetric force field reduce to 
\begin{eqnarray}
 \frac{dv}{dt} &=& \frac{q}{m} E(r,t) \label{eq_dvdt}\\
 \frac{dr}{dt} &=& v \label{eq_drdt}
\end{eqnarray}
where $E(r,t)$ is the radial electric field experienced by 
a particle at distance $r$ from the cloud's centre. 
We shall verify a posteriori that neglecting the centrifugal term 
$L^2/(m^2 r^3)$, which normally appears on the rhs of equation (\ref{eq_dvdt}), 
is justified by the fact that the field at the particle's position 
decreases asymptotically as $r(t)^{-2}$ (cf Section \ref{sec_efield}), which 
is slower than the $r(t)^{-3}$ dependence from the centrifugal term.   
   
Given the spherically symmetric field
$E(r,t)$ assumed in (\ref{eq_dvdt}) the particles must be interpreted as
infinitely thin spherical shells rather than point particles. This approximation
is justified as long as the number of particles within a given
spherical shell is large with respect to unity, i.e.
for radial distances $\gg (3/4\pi n)^{1/3}$ at time $t=0$.
Strict spherical symmetry reduces the original three dimensional system 
to a one dimensional system which can be treated much faster on a computer.
The main drawback is an unrealistic description of the
central part of the cloud which does not really matter as
the small (and continuously decreasing) number of particles living in this
region makes these particles statistically irrelevant anyway. 
Equations (\ref{eq_dvdt}) and (\ref{eq_drdt}) must be supplemented
by an equation for the electric field which   
for a distribution of $N$ thin spherical shells of radius
$r_k(t)$ and charges $q_k$ is simply
\begin{equation}
 E(r,t) = \frac{Q(r,t)}{4\pi\varepsilon_0 r^2}, \;\; {\rm with}\;Q(r,t) =
\displaystyle{\sum_{r_k(t)<r} q_k}. \label{eq_E}
\end{equation}

\section{Setup and parameters of a selected simulation\label{sec_setup}}

  \begin{figure}[h!]
  \center{\resizebox{!}{0.6\textwidth}
  {\includegraphics{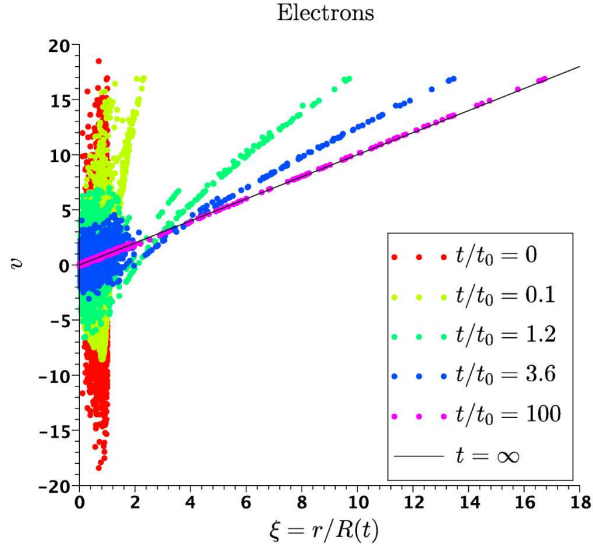}}}
  \caption{Snapshots of a subset of 3000 electrons in phase
space. Initially, at time $t=0$, the electrons are distributed uniformly
within a spherical shell $0.1<\xi<1$ with radial velocities following a
Maxwell-Boltzmann distribution with thermal velocity $v_{\rm e}=8.34$
(cf equation (\ref{eq_maxw}).}
  \label{fig1}
  \end{figure}

  \begin{figure}[h!]
  \center{\resizebox{!}{0.6\textwidth}
  {\includegraphics{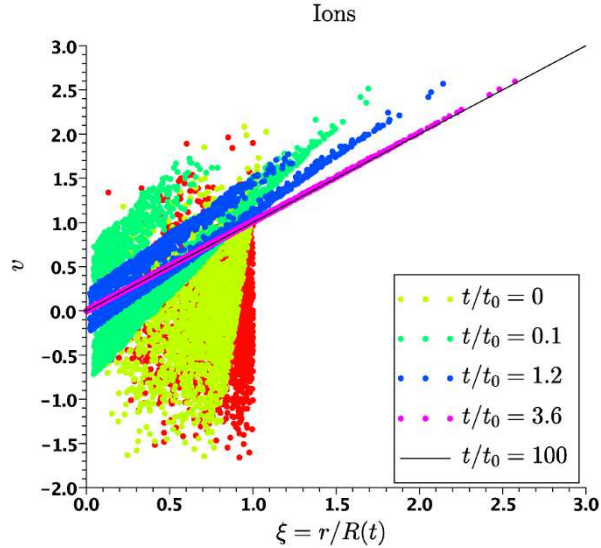}}}
  \caption{Snapshots  of a subset of 3000 ions in phase space.
Initially, at time $t=0$, the ions are distributed uniformly within a
spherical shell $0.1<\xi<1$ with radial velocities following a
Maxwell-Boltzmann distribution with thermal velocity $v_{\rm i}=0.834$
(cf equation (\ref{eq_maxw}).}
  \label{fig2}
  \end{figure}
Figures \ref{fig1} and \ref{fig2} show
the motion in phase space of a fraction of electrons and ions
from a numerical simulation of a spherical
cloud expanding into vacuum.
Positions and velocities of 
$N$ particles ($N/2$ electrons and $N/2$ ions)
are time advanced according to equations (\ref{eq_dvdt})
and (\ref{eq_drdt}) using a classical third order
leap frog integration scheme \cite{Quinlan_Tremaine_1990}.
The electric field is computed at every time
step using the updated particles' positions and
the field equation (\ref{eq_E}).

The initial conditions consist in $N=80000$
particles uniformly distributed within the
spherical shell $r_{\rm min}<r< R_0$ corresponding to $\Lambda = 0.0538$. 
Thus, even though the simulated number 
of particles $N$ is much smaller than in a typical dust impact
generated plasma cloud, it is still large enough for the key
parameter $\Lambda$ to be much smaller than unity so that the expansion is
quasi-neutral. 

The inner sphere $r<r_{\rm min}$ cannot be penetrated by particles
and is merely there to avoid the divergence of the Coulomb potential
for $r \rightarrow 0$ when particles (actually thin spherical shells) approach
the central region. In practice we chose $r_{\rm min}=0.1 R_0$,
which is both sufficiently small to
minimise its influence on the overall system's evolution and sufficiently 
large with respect to the strong interaction radius 
$r_{\rm s}$ to rule out binary collisions and self-charge effects. 

In the following, if not otherwise stated, we normalise charge to the elementary charge e,
mass to the electron mass $m_{\rm e}$, length to $r_{\rm min}$, electric field to
$E_{\rm n}\equiv e/(4\pi\varepsilon_0 r_{\rm min}^2)$, velocities to $v_{\rm
n}\equiv e/(m_{\rm e} 4\pi\varepsilon_0 r_{\rm min})^{1/2}$, 
time intervals to $t_{\rm n} \equiv r_{\rm min}/v_{\rm n}$ 
and temperatures to $T_{\rm n}\equiv m_{\rm e} v_{\rm n}^2$.
With these normalisation, and by consistently normalising density to $r_{\rm min}^3$,  
the Debye length (\ref{eq_ld}) reads $l_{\rm D}=(T/4\pi n)^{1/2}$, the
mean free path (\ref{eq_mfp}) $l_{\rm e}=T^2/(\pi n \lambda)$ and the electric field of a point
charge $Q$ becomes $E=Q/r^2$.  
We then set the initial temperature, for both electrons and ions,
to $T_0=34.76$ and the cloud radius to $R_0=10$. The resulting Coulomb logarithm
is then $\lambda=4.03$ and according to (\ref{eq_mfp}), the mean
free path $l_{\rm e}$ is equal to $R_0=10$ as postulated.
In code units the thermal velocity of the electrons is $v_{\rm
e}=(2T_0)^{1/2}=8.34$ and the strong interaction radius
$r_{\rm s}=1/3T_0=9.6\:10^{-3}$ which, as required, is much smaller
than both $R_0=10$ and $r_{\rm min}=1$. 

For convenience in Figure \ref{fig1} and in all subsequent figures
we use normalised positions $\xi\equiv r/R(t)$ with the temporal variation 
of the scale length defined by $R(t)\equiv R_0(1+t/t_0)$. We choose to set 
the arbitrary constant $t_0=10$ in order to have $d{R}/dt=1$.  
The ion to electron mass ratio is set to $m_{\rm i}/m_{\rm e}=100$ 
so that $t_0$ does actually turns out to be of the order of 
the ion sound crossing time $R_0/(3T/m_{\rm i})^{1/2}=9.8$, a 
characteristic time for the initial system.
As already stated, at $t=0$ particles are uniformly distributed
within the spherical shell $0.1< \xi \leq 1$ following   
Maxwell-Boltzmann velocity distributions for both ions and electrons: 
\begin{equation}
 f_j(v)= \frac{n_0}{\pi^{1/2}
\:v_j}\:e^{-\displaystyle{(v/v_j)^2}}\label{eq_maxw}
\end{equation}
where $v_j=(2T_0/m_j)^{1/2}$ is the thermal velocity of the
corresponding species $\;j={\{\rm e,i\}}$.

\section{Asymptotic evolution, theoretical background\label{sec_asymptotic}}

The particle trajectories shown in 
Figures \ref{fig1} and \ref{fig2} illustrate two key 
aspects of the expansion which will be discussed in sections \ref{sec_traj}
and \ref{sec_phase}.
First, as $t\rightarrow \infty$, all trajectories are seen to 
collapse towards the curve $v=\xi$. As a consequence the temperature
at a given position $\xi$ is seen to decrease with time as the
particles velocities appear to be less and less scattered as time progresses.
Second, whereas ions rapidly line up in a
structureless ribbon along the $v=\xi$ curve, electrons 
converge towards a more complex structure, also aligned on 
the $v=\xi$ curve, but with a bulging of the ribbon in the region  $\xi\lesssim
2$.
As we shall see below the bulging is due to the
bouncing motion of electrons trapped in an electrostatic potential well.

\subsection{Asymptotic convergence of particle 
trajectories\label{sec_traj}}

In this section we show that for
$t\rightarrow\infty$ all particle trajectories must end up
onto the $v=\xi$ curve provided the electric field
decays sufficiently fast everywhere in the system.
To this end we Taylor expand the asymptotic 
evolution of a particle velocity in terms of the
small parameter $\nu=t_1/t\ll 1$, i.e. $v(\nu)= v(0)+\nu(\partial
v/\partial\nu)_{\nu=0}$, where $t_1\gg t_0$ is just an arbitrary
finite time level. From the
equation of motion (\ref{eq_dvdt}) we obtain the asymptotic
evolution of a particle's velocity  
\begin{equation}
 v(t\gg t_1)=  v_{\infty} - \frac{t^2}{t_1}
\frac{qE(\xi,t)}{m}
\bigg|_{t\rightarrow \infty} \left( \frac{t_1}{t} \right) = v_{\infty}
- \frac{qE(\xi,t)t}{m}\label{eq_v_asympt}
\end{equation}
where $v_{\infty}\equiv v(t=\infty)$.  
The asymptotic evolution of the particle's position is formally
obtained by integrating (\ref{eq_drdt}), i.e.
$r(t)=r_1+\int_{t_1}^{t} v(\tau)d\tau$. Using $v(t)$ from equation
(\ref{eq_v_asympt}), and for $t\gg t_1$ one obtains
\begin{equation}
 \xi(t) = \frac{r_1}{t}+v_\infty - \frac{1}{t}\int_{t_1}^{t} d\tau
\frac{qE(\xi,\tau)\tau}{m} \label{eq_xi_asympt}.
\end{equation}
The last term on the right hand side of equation (\ref{eq_xi_asympt}) vanishes
for $t\rightarrow\infty$ provided $E$ at position $\xi$
decays faster than $t^{-1}$ in which case $\xi_\infty \equiv
\xi(t\rightarrow\infty)
=v_{\infty}$, confirming that the end point of a particle's trajectory
lies on the $\xi=v$ curve.  
For a time dependence of the electric field $E\propto t^\alpha$
it is possible to compute the slope of a particle's trajectory
in the phase space directly from equations (\ref{eq_v_asympt}) and
(\ref{eq_xi_asympt}). Indeed, assuming that for $t\rightarrow \infty$ the
variation of the electric field at a given particle position is
due primarily to the time dependence of $E$ rather than to
the particle's motion, one obtains
$(v-v_\infty)/(\xi -\xi_\infty) = 2+\alpha$.
Thus, for $\alpha = -2$, corresponding to the final, self-similar, evolution
of our system (see Figure \ref{fig8}), $(v-v_\infty)/(\xi -\xi_\infty)=0$, i.e.
trajectories approach the $\xi=v$ curve on horizontal trajectories with $v={\rm
const}$. 
In the particular case where $E(\xi,t)=E_1(\xi) t_1^2/t^2$ 
(which applies to the simulation for $t/t_0 \gg 1$) 
equations (\ref{eq_v_asympt}) and (\ref{eq_xi_asympt}) reduce to
\begin{eqnarray}
   v &=&  v_{\infty} - \frac{qE_1(\xi)}{m}\:\frac{t_1}{t}\label{eq_v_inf}\\
 \xi &=& \xi_\infty + \frac{r_1}{t}  -
\frac{qE_1(\xi)}{m}\:\frac{t_1}{t}\: \ln\left(\frac{t}{t_1}\right) \simeq
\xi_\infty - \frac{qE_1(\xi)}{m}\:\frac{t_1}{t}\:
\ln\left(\frac{t}{t_1}\right)\label{eq_xi_inf}
\end{eqnarray}
Equation (\ref{eq_xi_inf}) shows that for sufficiently late times
$|\xi(t)-\xi_\infty|/\xi_\infty\ll 1$ confirming that
the variation of the electric field at particle's position
is asymptotically dominated by the field decay and not by
the particle's motion. 
The interesting point about equation (\ref{eq_xi_inf}) is that it shows that
for $t\rightarrow \infty$ (which allows neglecting the $r_1/t$ term) particles
approach their final position $\xi_\infty$ from the left or the right depending
on the sign of $qE_1$. Thus, in an overall positive electric field, 
which is indeed the case for the expanding cloud problem
at hand (see Figure \ref{fig8}),
ions approach their final position from the left in $(\xi,v)$
space while electrons approach their final position from the right.
Thus, ions (electrons) which are initially on the right (left)
of the $v=\xi$ curve will first cross the $v=\xi$ curve before converging
towards their asymptotic position on horizontal $v\approx{\rm const}$ 
trajectories. This behaviour is already visible on the
early phase of the expansion shown on figures \ref{fig3} and \ref{fig4}.

\subsection{Shrinking of the volume occupied by particles in
$(\xi,v)$  space
\label{sec_phase}}

The shrinking of the phase space volume occupied by the particles in the
$(\xi,v)$ phase space is merely the  consequence of the
time dependence of the scaling length $R(t)$. 
The equations of motion for an individual particle (\ref{eq_dvdt}) and
(\ref{eq_drdt}) deriving from the general Hamiltonian 
of the system  
\begin{equation}
  H(r_1,...,r_N,p_1,...,p_N,t)= \sum_{j=1}^N p_j^2/2m_j + q_j\phi(r_j,t)
\end{equation}
where $p_j\equiv m_jv_j$ and $-\partial\phi/\partial r_j= E(r_j,t)$, 
it follows that any volume $\Gamma=\int dvdr$ must be conserved along particle
trajectories in $(v,r)$ space.
Thus, $\Gamma = R(t)\int dvd\xi = const$ with the
consequence that $\int dvd\xi \propto R(t)^{-1} =1/t$, i.e. the
volume covered by the particles in the $(\xi,v)$ phase space shrinks
in time as $1/t$.
On the long term all particles must end up aligned on the $v=\xi$ curve
with the spatial distribution of the charges being a function of the 
initial conditions, i.e. on the dimensionless parameter
$\Lambda(N)\equiv\lambda_D/R_0$ only.  

\section{Trapping, bouncing and freezing of particle
trajectories\label{sec_trapp}}

  \begin{figure}[h!]
  \center{\resizebox{!}{0.8\textwidth}
  {\includegraphics{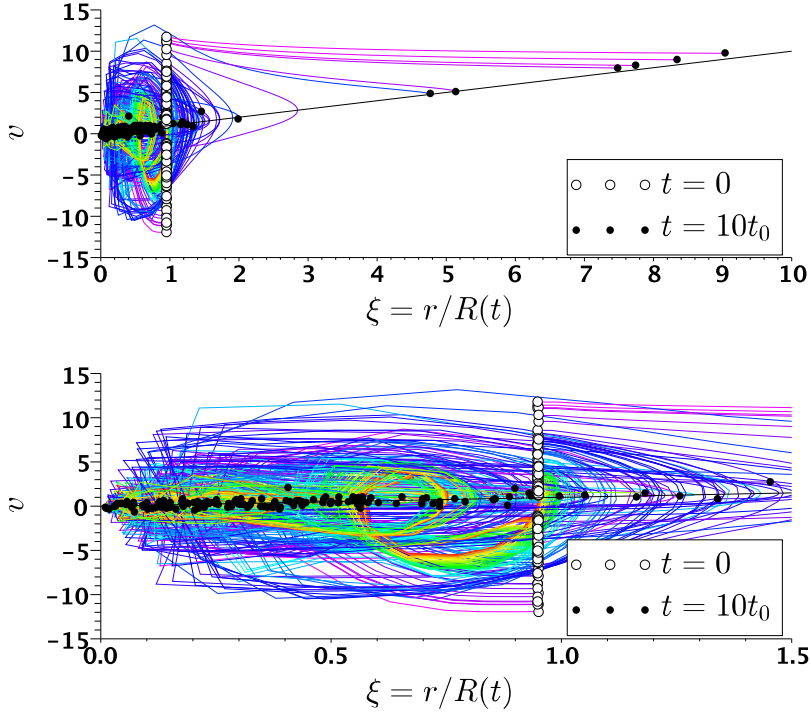}}}
  \caption{Sample trajectories of 102 electrons initially located in the
range $0.949< \xi <0.95$ with velocities $-12<v<12$. The bottom panel is
a zoom of the top panel.}
  \label{fig3}
  \end{figure}

  \begin{figure}[h!]
  \center{\resizebox{!}{0.6\textwidth}
  {\includegraphics{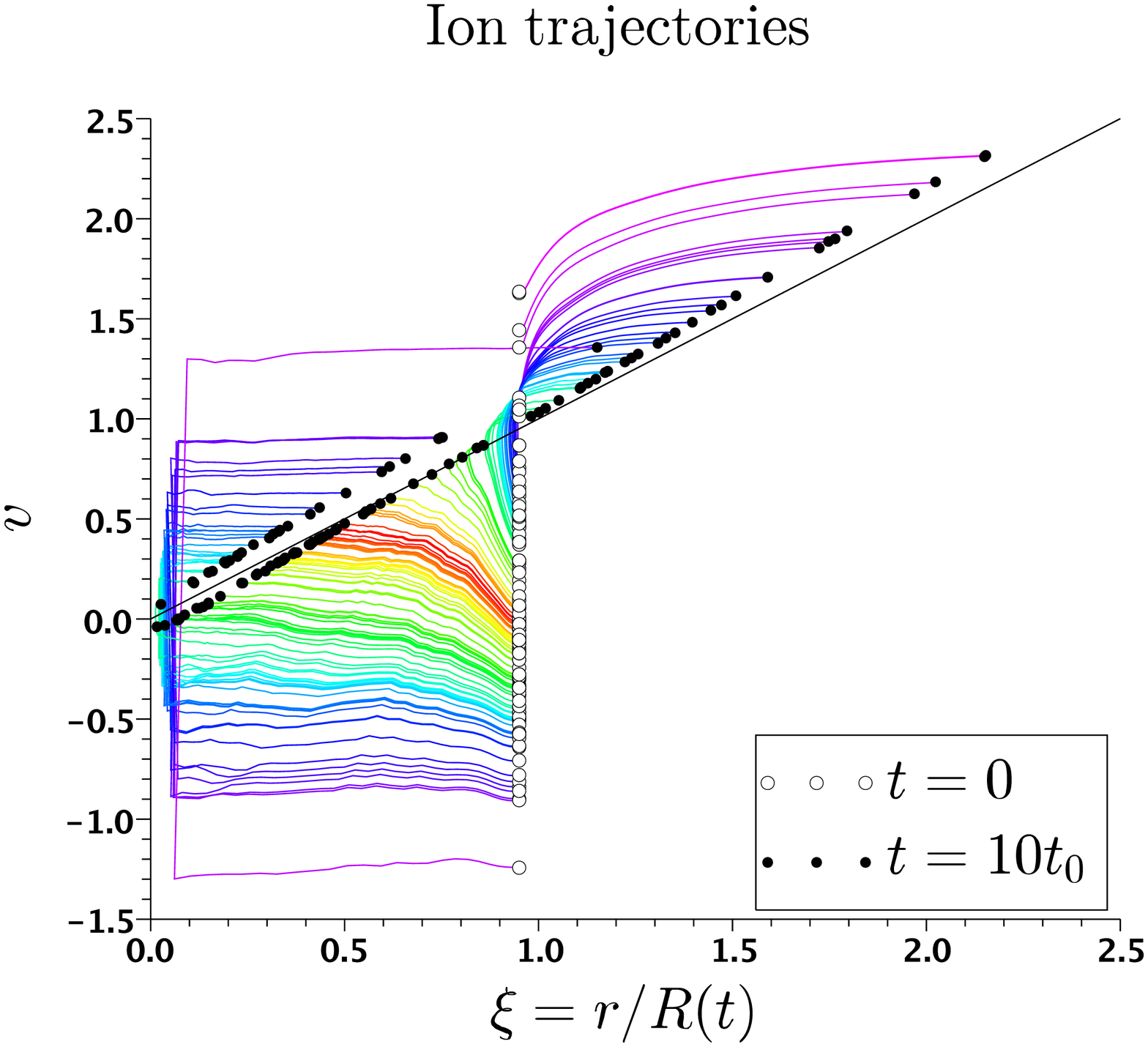}}}
  \caption{Sample trajectories of 116 ions initially located in the
range $0.949< \xi <0.95$ with velocities $-2<v<2$.}
  \label{fig4}
  \end{figure}
\begin{figure}[h!]
  \center{\resizebox{!}{0.6\textwidth}
  {\includegraphics{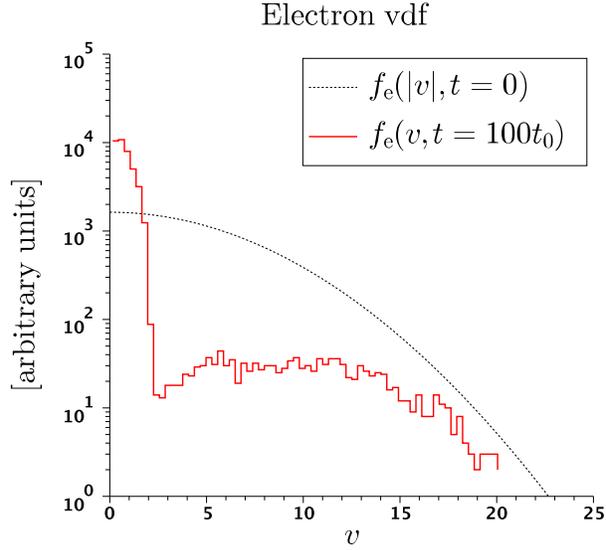}}}
  \caption{Electron velocity distribution function $f_{\rm e}(v)$ at the end of
the simulation at $t=100t_0$. Plotted as a reference, the
initial distribution of the absolute value of the radial velocities $f_{\rm
e}(|v|)$.}
  \label{fig5}
  \end{figure}

Figures \ref{fig3} and \ref{fig4} show characteristic trajectories of selected 
electrons and ions in ($\xi,v$) phase space. From the figures it is
immediately apparent that both species behave in a radically different way.
Ions follow rather dull trajectories and are either accelerated outwards
(in particular the outermost ones) or move at approximately constant velocity.
The fastest electrons (in general the ones at largest radial
distance $\xi$)
are seen to steadily reduce their outflow velocity in the attractive 
field of the positively charged interior of the cloud. 
However, electrons, with sufficiently low initial energy (the ones with
end velocity $v\lesssim 5$), do cross the $v=\xi$
curve and eventually bounce within an electrostatic trap.
In order to understand the trajectories in the $(\xi,v)$ space
it may be useful to rewrite the equations of motion
(\ref{eq_dvdt}) and (\ref{eq_drdt}) by setting $v=v(\xi(t))$, viz
\begin{eqnarray}
 (v-\xi) \frac{\partial v}{\partial \xi} &=&\frac{q}{m} E(\xi,t)\: t
 \label{eq_dvdxi}\\
 \dot{\xi} &=& \frac{v-\xi}{t}\label{eq_dxidt}
\end{eqnarray}
where we have used $dv/dt = \dot\xi \partial v/\partial \xi$ with
$\xi=r/R(t)=r/t$ and $\dot{R}=1$. Equation (\ref{eq_dvdxi}) shows that 
when a trajectory crosses the $\xi=v$ it satisfies the condition $\partial
v/\partial \xi=\infty$ (cf Figures \ref{fig3}
and \ref{fig4}) unless the electric field $E(\xi,t)=0$. In this
particular case the particle's velocity is constant $v=v_\infty$ and 
(\ref{eq_dxidt}) shows that it takes an infinite time for the particle 
to reach the $v=\xi$ curve as $\xi(t)-\xi(\infty) \propto 1/t$.
Multiple reflections are associated with an equal number
of crossings of the $v=\xi$ curve, from top to bottom for an
inward directed force and from bottom to top for an outward directed force.
Note that particles approaching the centre $\xi=0$
make an artificial reflection there as their radial velocity must change from
negative to positive. Given that in the simulations particles are not
allowed to approach the centre at a distance less than $\xi_{\rm min}(t)=0.1 R_0/R(t)$  
reflection does effectively occur at $\xi_{\min}$ instead.
Electrons bouncing back and forth in an expanding potential well
can be efficiently cooled by first order Fermi deceleration.
Now, despite the fact that both, bouncing and non bouncing electrons lose
kinetic
energy, the phase space volume occupied by the two
populations evolves differently in time.
Thus, whereas the velocity difference $\Delta v=|v_2-v_1|$
between two electrons with initial velocities $v_2$ and $v_1$
increases in time for $v_1,v_2\gtrsim 10$, 
the opposite is true for the trapped (and eventually bouncing)
electrons with initial velocities  $v_1,v_2\lesssim 10$ (see Figure \ref{fig3}).
Loosely speaking, trapped electrons 
contribute to rise the particle concentration $f_j/\Delta v$
in velocity space while non trapped electrons, and basically all ions, 
contribute to reduce the particle concentration in velocity space.
Both effects are visible in Figure \ref{fig5} which shows that
$f_{\rm e}(v)$ substantially grows over its initial value
for $v\lesssim 1.5$ where Fermi deceleration occurs.
For $v\gtrsim 1.5$ the electron trajectories in velocity space are divergent
and $f_{\rm e}(v)$ falls below its initial value, instead. 

\begin{figure}[h!]
  \center{\resizebox{!}{0.5\textwidth}
  {\includegraphics{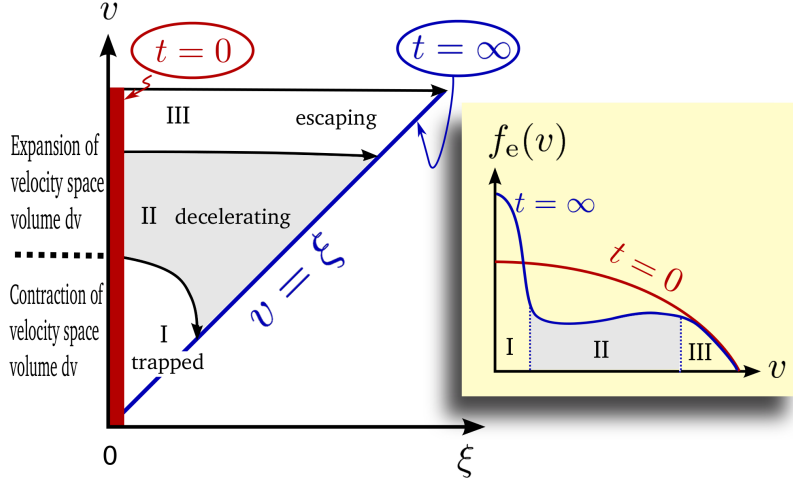}}}
  \caption{Schematic illustration of the electrons' evolution
in the $(\xi,v)$ phase space (left graph) and of the electron velocity
distribution function $f_{\rm e}(v)$ (right graph). }
  \label{fig6}
  \end{figure}

A schematic representation of the electron dynamics in
phase space and the corresponding evolution of their velocity distribution
function $f_{\rm e}(t)$ summarising the discussion of this section is shown in
Figure \ref{fig6}.

\subsection{Freezing of the bouncing motion}

Figure \ref{fig3} shows that some electrons have time to bounce several times
during the simulation 
before their motion becomes frozen onto the $v=\xi$ curve. Others, for example
the ones with initial velocity $v\approx 10$ and final velocity  near $v\approx
3$
are unable to perform a full bounce period during the time of the simulation.
The reason is that the bouncing period $t_{\rm b}$ for a particle
of mass $m$ and charge $q$ moving in
the potential $Q/r$ of a point charge $Q$ is of the order $t_{\rm b} \sim 2\pi (m
r^3/qQ)^{1/2}$ (Kepler's third law) so that $t_{\rm b} \propto \xi^{3/2} R^{3/2}
\propto
t^{3/2}$ grows faster than the expansion time $t_{\rm exp}= r/v = \xi R/v$.
Thus, as the bouncing time to expansion time ratio $t_{\rm b}/t_{\rm exp} \sim
2\pi(v^2 m R_0/qQ)^{1/2} \xi^{1/2} (R/R_0)^{1/2}$ grows as $t^{1/2}$
the bouncing motion for a given particle will become frozen  
as soon as $t_{\rm b}(t,\xi,Q)/t$ exceeds a value of order unity.
Electrons for which $t_{\rm b}/t_{\rm exp} \approx 1$ may just be a
able to bounce once but may still lose a significant fraction
of their initial kinetic energy if their total energy (kinetic + potential) is
small and negative.
A particle hitting a wall moving at velocity $w$ is slowed down
by $\delta v =-2w$ or $\delta v = 2(w-v)$ depending
on whether its initial velocity $v$ is in the range $v\ge 2w$ or
$2w>v>w$, respectively. Given that the  
expansion velocity of the electrostatic walls
is rather slow  compared to the electron thermal velocity $v_{\rm e}=8.3$ 
(the peak of the electrostatic profile in Figure \ref{fig8}
at $\xi\gtrsim 2$ moves outwards at a velocity $\lesssim 2$),
2 or 3 reflections are required to slow down a
thermal electron below the critical velocity $v^* \approx 2$
which separates 
bouncing and non bouncing electrons (see Figures \ref{fig5} and \ref{fig6}).
The maximum initial velocity $v^*$ of a trapped
electron can be estimated by equating the bouncing period $t_{\rm b}$
and the expansion time $t_{\rm exp}=\xi R/v$ 
\begin{equation}
  v^* \approx
\frac{R^{-1/2}}{2\pi}\left[\frac{Q(\xi)}{\xi}\right]^{1/2}\label{eq_v_star}.
\end{equation}
Equation (\ref{eq_v_star}) may appear rather useless,  
$Q(\xi)$ being an unknown function of $\xi$.
However, we expect the right hand side of equation
(\ref{eq_v_star}) to be largest after a short time of the
order $\lambda_{\rm D}/v_{\rm e}=0.091$ (the inverse electron
plasma frequency) from the beginning of the
simulation. Thus, an a priori estimate is possible if we assume
that $Q$ corresponds to the number of electrons in the outer Debye
shell of the initial cloud,
i.e. $Q\approx 4\pi \lambda_{\rm D} R_0^2 n_{\rm e}=1.5N
\Lambda$.
In the present case $N=80000$ and  $\Lambda=0.0538$ so that $Q\approx
6456$. With these numbers and by setting
$R=R_0=10$ and $\xi=1$ in equation (\ref{eq_v_star}), one obtains $v^* \approx
4$
which appears to be a fair estimate of the upper limit for the initial
velocity of trapped electrons (see Figure \ref{fig3}).

\subsection{Ion distribution}

\begin{figure}[h!]
  \center{\resizebox{!}{0.6\textwidth}
  {\includegraphics{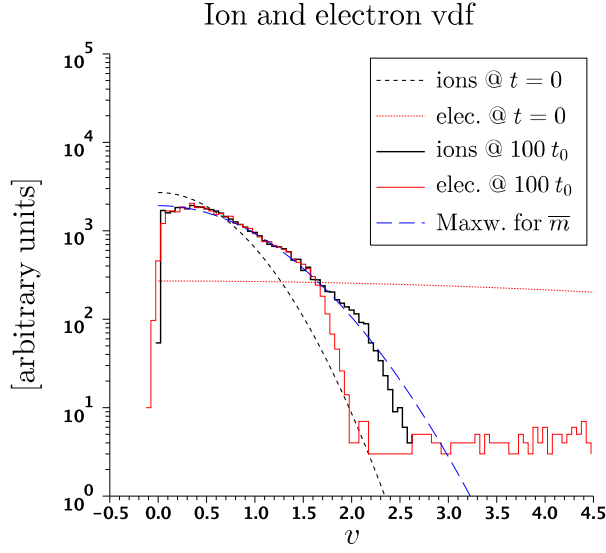}}}
  \caption{Ion and electron velocity distributions $f_k(v)$ at the end
of the simulation at $t=100t_0$. Also plotted are the initial distributions
$f_k(|v|,0)$ and the Maxwellian $N(\pi v_{\overline{m}})^{-1/2}
\exp[-(v/v_{\overline{m}})^2]$ where $v_{\overline{m}}
\equiv (2 T_0/\overline{m})^{1/2}=1.17$ is the thermal velocity based on the
initial
temperature $T_0=34.76$ and a particle of mass
$\overline{m}\equiv (m_{\rm i}+m_{\rm e})/2=50.5$.}
  \label{fig7}
  \end{figure}

The asymptotic ion velocity distribution is shown in Figure \ref{fig7}.
The figure clearly shows that the ion distribution closely follows the
electron distribution for $v\lesssim 1.7$ (roughly twice the
ion thermal velocity $(2T_0/m_{\rm i})^{1/2}=0.83$)
while at higher velocities,
up to the velocity of the fastest ions in the simulation $v \approx 2.5$, 
the ion density is in excess over the electron density.
On the other hand, for $v\lesssim 0.5$ the asymptotic distribution
$f_{\rm i}(v,\infty)$ falls below its
initial value $f_{\rm i}(v,0)$ while the opposite occurs
for  $v\gtrsim 0.5$.
In principle, given that the mobile electrons tend to escape
from the cloud, all ions should be accelerated outwards by the
positive charge $Q(\xi)$ of the remnant (see bottom panel
of Figure \ref{fig9}) and its associated, 
outwards directed electric field (see Figure \ref{fig8}).
The outwards directed force
should produce a displacement towards larger radial velocities
of the original velocity distribution
$f_{\rm i}(|v|,0)$ with $f_{\rm i}(|v|,t)=0$
below some minimum velocity. The displacement of $f_{\rm i}$ towards
higher radial velocity is visible in \ref{fig7} for the 
fastest ions with end velocities $v\gtrsim 0.5$. However, no region with $f_{\rm
i}=0$ 
is visible at low velocities, though. The reason is that the 
slow ions are kept in the inner part of the cloud
by the inwards falling, Fermi decelerated, electrons. The coupling
between ions and cold electrons is so efficient there
that the electric field asymptotically goes to zero
for $\xi\lesssim 1.5$ (see Figure \ref{fig8}).
Figure \ref{fig7} also shows that 
up to $v \approx 1.7$ both electron and ion velocity 
distributions are conveniently fitted by a single Maxwellian
distribution with thermal speed $v_{\overline{m}}=(2T_0/\overline{\rm
m})^{1/2}=1.17$
where $\overline{m}\equiv 0.5(m_{\rm i}+m_{\rm e})=50.5$
is the average particle mass. If we assume
that the electron distribution is a Maxwellian sharply cut
at an upper velocity $v^*$, 
such that the missing electrons are those in the outer Debye shell
of the initial sphere  $1.5\Lambda N = 6456$, we obtain the estimate
${\rm Erf}(v^* /v_{\overline{m}})=1-1.5\Lambda$
(where ${\rm Erf}(x) \equiv 2\pi^{-1/2}\int_0^x ds\:
\exp(-s^2)$ is the error function) which gives $v^* = 1.44$.
Obviously, $1.5\Lambda N$ is an overestimate of the number of
electrons in the electron precursor. From Figure \ref{fig9} we take that 
this number is roughly 5 times smaller than $1.5\Lambda N$
which allows for a more realistic estimate
${\rm Erf}({v^*}/v_{\overline{m}})=1-1.5\Lambda /5$, i.e. $v^* =
1.99$.

\section{Charge distribution and electric field\label{sec_efield}}

\begin{figure}[h!]
  \center{\resizebox{!}{0.6\textwidth}
  {\includegraphics{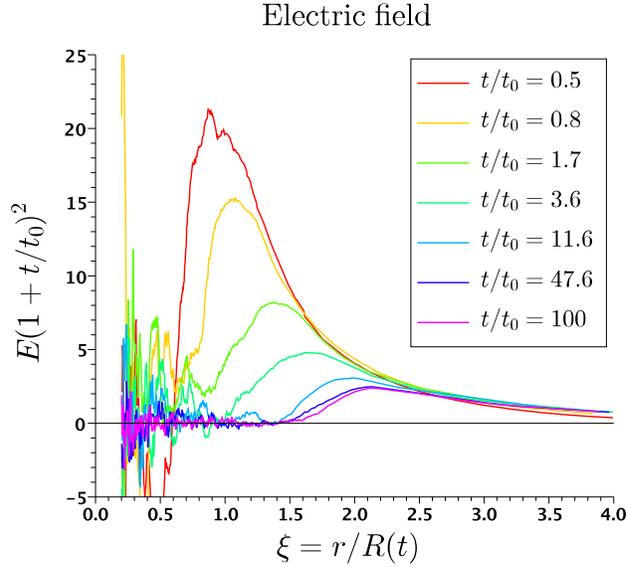}}}
  \caption{Temporal evolution of the electric field profile. For $t\gtrsim 30$
the spatial profile is essentially frozen while the amplitude declines $\propto
t^{-2}$.}
  \label{fig8}
  \end{figure}

\begin{figure}[h!]
  \center{\resizebox{!}{0.7\textwidth}
  {\includegraphics{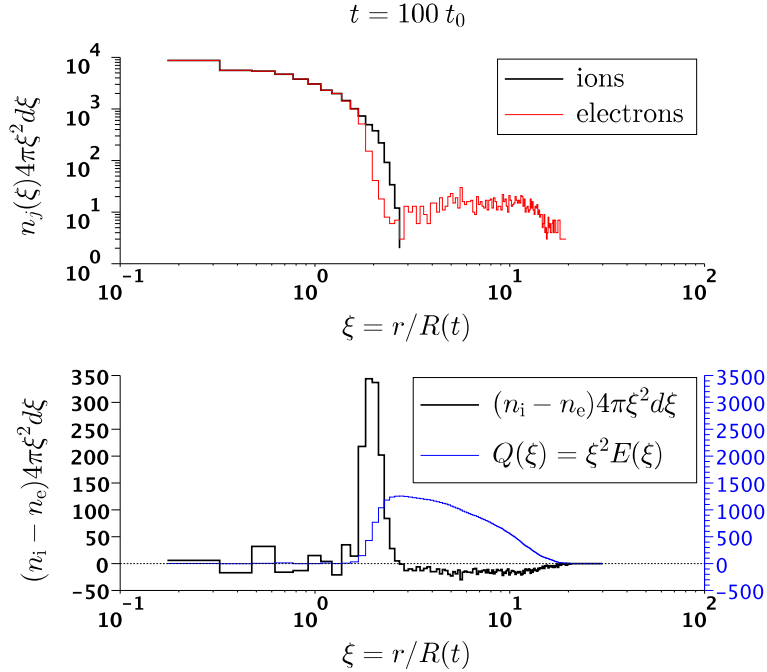}}}
  \caption{Radial linear density for ions and electrons (top panel).
The bottom panel shows the linear charge density (thick black curve and 
right scale) in units of the positive elementary charge $|e|$. The thin blue
curve corresponds to the 
net charge $Q(\xi)$ contained within the the spherical shell of radius $\xi$.}
  \label{fig9}
  \end{figure}

The spatial and temporal structure of the electric field is shown in  Figure
\ref{fig8}.
As expected, after a dynamic initial phase, when particles are close 
to their asymptotic position in the $(\xi,v)$ space, the electric field
intensity decays as $E({\xi,t})
\propto t^{-2}$. The spatial structure $E(\xi,\infty)$ is characterised by
a central region $\xi\lesssim 1$ where, apart from fluctuations due to the
small number of particles , the field intensity
is essentially zero, corresponding to the region where the electron and ion
density are equal (see Figure \ref{fig9}).
For $\xi\gtrsim 1.6$ the field intensity rapidly rises towards the maximum
$E_{\max}(1+t/t_0)^2\approx 2.5$ at 
$\xi=2.1$, followed by a gentler negative slope over a much larger
spatial scale. The large scale is associated with the scale of
the electron precursor,
which is obviously forged by the initial electron velocity
distribution, and scales as
$\xi_{\rm fall}\approx v_{\rm e}t_0/R_0=8.34$.
The shorter scale of the rising part of the
field profile is forged by the ions and scales as
$\xi_{\rm rise}\approx v_{\rm i}t_0/R_0=0.83$.

The charge distribution at the end of the simulation at $t=100 t_0$ is shown in
the bottom panel of Figure \ref{fig9} where, again, 
$Q(\xi)$ represents the charge within
the sphere of radius $\xi$. The maximum $Q_{\rm max}\approx 1292$ is
reached for $\xi\approx 2.7$. Thus $Q_{\rm max} \approx 0.3N\Lambda$, 
corresponding to roughly $1/5$ of the positive charge contained in the
outermost Debye shell of the cloud at $t=t_0$. 
On the other hand the
charge at the position of the electric field maximum is 
$Q(\xi=2.1)\approx 1053\approx 0.245N\Lambda $ giving a maximum field intensity
$E_{\max}(t/t_0)^2 \approx 0.245N\Lambda/(2.1 R_0)^2 \approx 2.4$ as 
confirmed by the latest profiles of $E$ shown in Figure \ref{fig8}. 
Changing to SI units we obtain a more useful expression 
\begin{equation}
 E_{\max,{\rm SI}} = 8\:10^{-11}\frac{N\Lambda}{R^{2}}\label{eq_emax}
\end{equation}
where $E_{\max,{\rm SI}}$ is expressed in V/m and $R$ in metres.

\section{Ion and electron fluid velocities\label{sec_velocities}}

  \begin{figure}[h!]
  \center{\resizebox{!}{0.7\textwidth}
  {\includegraphics{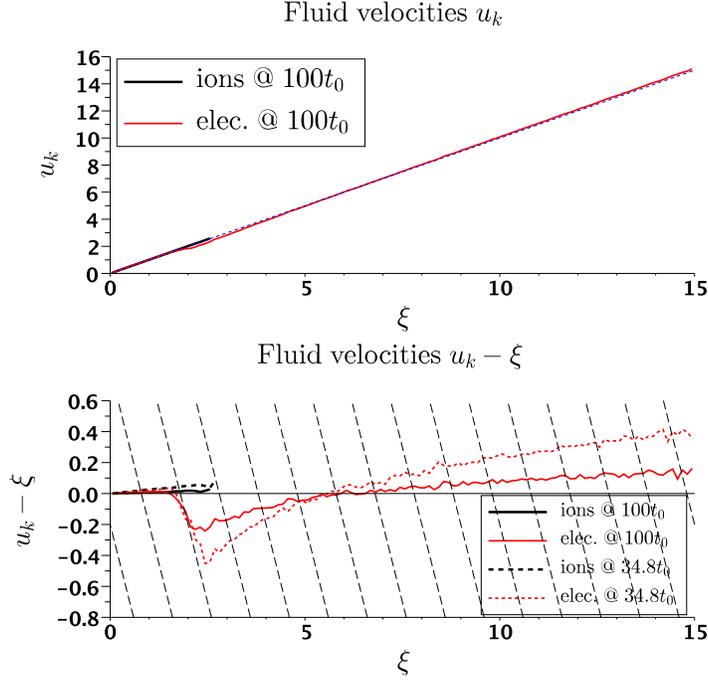}}}
  \caption{Fluid velocities $u_j(\xi)=\int_{-\infty}^{\infty} v\:
f_j(v,\xi)dv/n_j{\xi}$ for ions and electrons at the end of the simulation (top
panel) and $u_j-\xi$ for at two different times (bottom panel). The grid of
dashed lines in the bottom panel gives the trajectories $v={\rm const}$
individual particles are expected to follow in case of an electric field
declining as $t^{-2}$ (see Section \ref{sec_traj}).}
  \label{fig10}
  \end{figure}

Figure \ref{fig10} shows the spatial fluid velocity profiles $u_{\rm e,p}$
for both electron
and ions at the end of the simulation. Not surprisingly both populations have
velocities profiles close to the $v=\xi$ curve (top panel).
Plotting the fluid velocities $u_{\rm e,p} -\xi$ versus $\xi$ (bottom
panel) shows that while all ions and electrons located within $\xi \lesssim
1.7$ do closely follow the $v=\xi$ curve (meaning that they are already 
frozen) electrons with $1.7 \lesssim \xi \lesssim 6$ stay below
the $v=\xi$ curve. These electrons are still flowing
(falling)
inwards along the dashed lines, corresponding to
the $v={\rm const}$ trajectories predicted for a
$t^{-2}$ declining electric field (see Section \ref{sec_traj}).
The electron inflow velocity at $t=100 t_0$ peaks near $\xi\approx 2.3$
at about $10\%$ of the absolute ion fluid velocity. 
The ion velocity will not evolve significantly after $t=100 t_0$
as it is already well approximated by $u_{\rm i}=\xi$. The asymptotic
position of the electrons near $\xi\approx 2$ at $t=100 t_0$, 
will lie some $10\%$ closer to the centre of the cloud. 
This late displacement of the electrons
will not modify the final structure of the electric field significantly as the
density of these inflowing electron is substantially lower than
the ion density near $\xi=2$ (see Figure top panel in \ref{fig9}). 
On the other hand, electrons at $\xi(100 t_0) \gtrsim 3$ are too fast
for the ions to catch up with and 
constitute the final electron precursor.

\section{Application to the case of clouds formed by 
interplanetary nanoparticles impacts 
\label{sec_comparision}}

When dust particles travelling in the interplanetary space
hit a solid object at characteristic velocities of the
order of tens to hundreds km/s, a plasma cloud is generated 
at the impact point. The cloud is formed
due to the vaporisation and partial
ionisation of the dust particle itself and the target material. 
The subsequent expansion of the cloud is  
hemispherical rather then spherical as assumed in the above model. 
In the case of a non conducting and charge neutral target
the above results should not be modified in any substantial 
way. On the other hand, one should be extremely cautious 
when trying to interpret the electric signals measured 
on a spacecraft as spacecraft are generally 
positively charged due to electrons being stripped from 
their metallic surface through photoionisation by solar radiation. 
The  associated electric field (typically of the order of a few V/m) 
exceeds the cloud's internal electric field very early during
the expansion, causing stripping of most of the cloud's electrons 
\cite{Meyer-Vernet_Maksimovic_al_2009}. Not only is the expanding cloud subject
to charging so that the long term expansion is more like 
a Coulomb explosion\cite{Peano_Martins_al_2007,Beck_Pantellini_2009} rather than
a quasi-neutral expansion 
as described above, but the role of the photoelectrons, continually 
emitted and recollected by the spacecraft, must be taken into account 
when trying to interpret the voltage pulses measured on spacecraft antennae. 
Indeed, in the scenario described in appendix A of the Zaslawsky et al paper
\cite{Zaslavsky_Meyer-Vernet_al_2011}), the voltage pulses 
measured on individual antennae in conjunction with the impact of nano dusts on
the STEREO spacecraft are not a direct measure of the field within the post
impact expanding plasma cloud but are the consequence of the equilibrium 
photoelectron return current towards the part of the antenna within 
the plasma cloud being interrupted. The interruption of the 
photoelectron return current induces an accumulation of positive
charges on the antenna  which is then measured 
by on board detectors as a temporal variation 
of the potential between the antenna and the spacecraft.   

  \begin{figure}[h!]
  \center{\resizebox{!}{0.6\textwidth}
  {\includegraphics{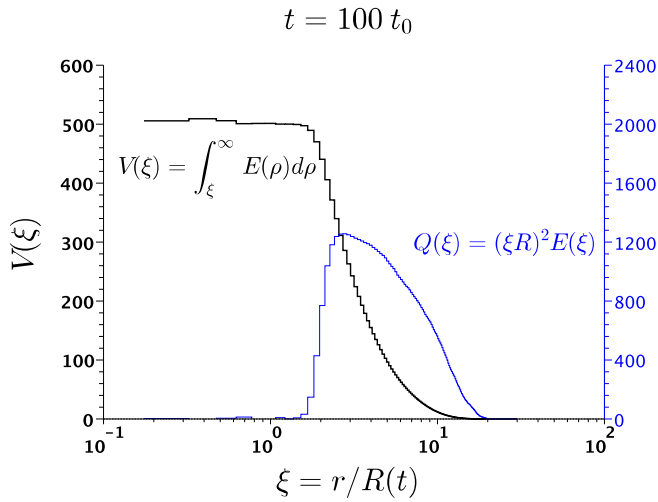}}}
  \caption{Electrostatic potential $V(\xi)$ and
total charge $Q(\xi)$ as in figure \ref{fig9}.
 }
  \label{fig11}
  \end{figure}

Let us estimate the electrostatic potential through a real cloud generated
by a nanoparticle hitting a target in  interplanetary space 
using the above model. As we shall see the intrinsic field of the cloud 
is too weak to account for the impact associated potential pulses 
observed on STEREO.  
 
The electrostatic potential through the simulated plasma cloud
is shown in Figure \ref{fig11}. The total drop in the 
electrostatic potential is $\Delta V \approx 550$
while, as already stated, the peak charge is 
$Q_{\rm max} \approx 1292$ (see bottom panel in Figure \ref{fig9}).
Assuming a linear relation between $\Delta V$ and $Q_{\rm max}$ 
we find  
\begin{equation}
\Delta V(N) \approx {Q_{\rm max} \over 2.35} \approx 0.128~N\Lambda~~~({\rm
simulation ~~units})\label{eq_deltav}.
\end{equation}
This relation can be interpreted as the electrostatic potential
at the surface of a sphere of radius $2.35 R$ enclosing a charge $Q_{\rm max}$.
Multiplication of Equation (\ref{eq_deltav}) by
$e/(4\pi \varepsilon_0 R)$ in SI units, leads to the
dimensional version of (\ref{eq_deltav}):
\begin{equation}
 \Delta V_{\rm SI}(N,R) \approx  1.84 ~10^{-10}\;{{N\Lambda} \over R}
\label{eq_deltav_SI}
\end{equation}
where $R$ is in meters and $V_{\rm SI}$ in Volts. 

Let us use equation (\ref{eq_deltav_SI}) to estimate
the voltage pulse $\Delta V$ due to the
impact of a $m_{\rm d}=10^{-20}\,$kg grain (size $\sim 10\,$nm)
travelling at $v=300\,$km/s. The number of electrons
and ions within the plasma cloud can be 
estimated via the semi-empirical formula $N/2 \approx 0.7 m_{\rm d}^{1.02}
v^{3.48}/e$ (see \cite{Meyer-Vernet_Maksimovic_al_2009,Hornung_Kissel_1994}),
where $[e]={\rm C}$,
$[m_{\rm d}]=$kg and  $[v]=$km/s. With these parameters
we obtain $N\approx 1.4~10^7$ and, from the relation between
$\Lambda$ and $N$ established in the discussion following equation
(\ref{eq_lambda}) we find $\Lambda(1.4~10^7) = 0.016$.
Plugging this value into  in equation (\ref{eq_deltav_SI})
leads to $\Delta V_{\rm SI}(1.4~10^7) \approx 4.1~10^{-5}/R$.
Thus, when the cloud's radius has grown to   
$R=50$cm, i.e. when its density has fallen to  
a value comparable to the interplanetary density,   
$\Delta V_{\rm SI} \approx 0.082 {\rm mV}$ only.
Noting that the effectively measured voltage is obtained by 
averaging over the whole antenna length, i.e. by multiplying the 
above voltage by $l/L$, where $L$ is the length of the antenna (6 m on STEREO)
and $l$ the length of the part of the antenna within the cloud, the 
voltage predicted by the model is far too week to be directly detectable
and in any case much too weak to induce the up to 100mV 
pulses observed on STEREO.   
Previous estimates of the voltage pulse associated 
with nano-size dust impacts were based on the assumption that
the charge separation within the expanding cloud is
total with $\sim N/2$ electrons in the
precursor\cite{Meyer-Vernet_Maksimovic_al_2009}.
The assumption, adopted here, that charge separation only occurs at the time
when the expanding cloud becomes collisionless, 
without influence of potential external fields,  
reduces the number of electrons in the precursor
to a much smaller number of order $0.3 N \Lambda$. 
Using equation (\ref{eq_emax}) we can estimate the maximum electric 
field within the above cloud to be of order $E_{\max,SI}\approx 1.8\;10^{-5}
R^{-2}$. Thus, when the cloud has reached $R=4.2 {\rm mm}$ the maximum field is
already down to 1V/m, i.e. comparable to the spacecraft's own field. Beyond
this size, the cloud will start loosing its electrons to the spacecraft. 
However, even in case of a complete charge separation  such that the 
electrostatic field outside the cloud is given by the Coulomb field $Q/r$, 
the latter is much too weak (at most a few mV) to account for the on board
measured voltage pulses associated with nanodust impacts.  
As we shall explain in more details in a forthcoming paper, but as already
briefly exposed in the appendix A of \cite{Zaslavsky_Meyer-Vernet_al_2011}, the
strong potential pulses measured on individual antennas on STEREO are due to
the perturbative effect of the expanding cloud on the photoelectrons surrounding
the antenna.

\section{Summary and conclusions\label{sec_conclusions}}
We have explored numerically the unconstrained spherically symmetric 
expansion of an initially uniform, overall neutral and at thermodynamic   
equilibrium cloud of immobile plasma. The initial temperature and density of the plasma       
are such that the cloud's radius equals the Fokker-Planck collisional mean free path
of a thermal electron, representing an admittedly crude model of an expanding 
cloud at the time it becomes collisionless.   
Consistently assuming that the ion and electron velocity distributions are
Maxwellians at the time of the collisional to collisionless transition,
it follows that the key parameter of the problem
$\Lambda\equiv\lambda_{\rm D}/R_0$  
can be written as a function of the total number of 
ions and electrons in the cloud only. 
Due to the $\Lambda \propto N^{-1/4}$ dependence (see equation
(\ref{eq_lambda}))
typical nano size grain impacts which are
expected to ignite plasma clouds with $N=\mathcal{O}(10^8)$,  
$\Lambda$ is always much smaller then unity, i.e. the collisionless
expansion is quasi-neutral.

During the initial phase of the collisionless regime 
most electrons (the less energetic ones)
lose nearly all of their kinetic energy
thorough Fermi deceleration in the expanding potential
(see Figure \ref{fig11}).
On the other hand the outermost ions
near the electric field maximum (see Figure \ref{fig8}), 
are accelerated outwards by the positive electric field.
The net effect is that ions and electrons do 
asymptotically tend towards having the same velocity distribution
up to a threshold of the order of the ion thermal velocity
(see Figure \ref{fig7}).
The fact that all particle trajectories
converge towards the $v=\xi$ curve as $t\rightarrow \infty$ 
implies that electron and ion fluid velocities end up being  
simple linear functions of the distance $r$ from the expansion centre.  

At late times the ion density profile is conveniently
described by $n_{\rm i}(r) \propto \exp[-(r/tv_{\overline{m}})^2]$
where $v_{\overline{m}}\equiv (2 T_0/\overline{m})^{1/2}$
is the thermal velocity based on the initial temperature $T_0$
and the mean mass $\overline{m}\equiv 0.5(m_{\rm i}+m_{\rm e})$.
Because of $\Lambda\ll 1$ the electron density $n_{\rm e}$
closely follows the ion density up to a distance
$r^*$ solution of the equation ${\rm Erf }(r^*/t v_{\overline{\rm
m}})\approx 1-0.3\Lambda N$. Beyond this point the 
electron density is flat up to a radial distance of the order 
of $v_{\rm e}t$ as observed in Vlasov simulations \cite{Manfredi_Mola_al_1993}.  
The electric field is essentially zero for $r\lesssim r^*$
(see Figure \ref{fig8}) but
rises towards a maximum on the ion length scale 
$r_{\rm rise}\approx v_{\rm i} t$
and decreases slowly on the electron precursor length scale
$r_{\rm fall}\approx v_{\rm e} t$.
At late times the maximum field intensity $E_{\rm max}$ is 
essentially nailed down  by the number of electrons in the outer shell of the
cloud of thickness $\lambda_{\rm
D}$. This number can be expressed in terms of the total number 
of particles $N$ and the dimensionless parameter $\Lambda\equiv\lambda_{\rm
D}/R_0$ to give $1.5 N\Lambda$, implying that the electric field 
intensity must be proportional to $N\Lambda/R^2$. In our
representative simulation, we find 
$E_{\max,{\rm SI}}\approx 8\:10^{-11} N\Lambda/R^2$ 
which we expect to hold as long as $N\gg1$ and provided 
the cloud has changed from collisional to collisionless 
during expansion. 
The electrostatic potential through the cloud has been found to 
be $\Delta V_{\rm SI}(N,R) \approx  1.84 ~10^{-10} N\Lambda/R$. 

The electrostatic potential difference between the cloud's centre 
and infinity predicted by the model is far too weak to be account  
for the voltage pulses,sometimes exceeding 100mV, observed on the S/WAVES TDS
detector on the STEREO spacecraft following a nano-sized dust particles impact.
One plausible reason for this discrepancy is that  
spacecraft are positively charged due 
to photoelectron emission through their sunlight exposed surfaces.
The resulting electric field, typically of the order of a few $V{\rm m}^{-1}$ 
at 1AU from the Sun, generally exceeds the
maximum field intensity within the plasma cloud  
before its dilution in the ambient plasma. The late evolution of the 
cloud is therefore dominated by the spacecraft field which 
strips most or all of the electrons from the cloud which then sees both its
charge and its internal electrostatic potential field increase by a factor of
order $\Lambda^{-1}\gg 1$. However, even in case of an unrealistically  
large nanodust impact generated cloud with some $N=10^8$ and  
with all electrons stripped-off, the total electrostatic potential difference 
would merely be a smallish $20{\rm mV}$
at the time of its maximum extension, when $R\approx 1{\rm m}$. Considering 
that the measured field is down by at least a factor $R/L$, where $L$ is the 
the total length of the antenna ($L=6{\rm m}$ on the Stereo spacecraft), one
must conclude that the on board measured fields are not a direct measure of the 
cloud's  intrinsic field. 
Indeed, recent findings by Zaslavsky et al \cite{Zaslavsky_Meyer-Vernet_al_2011}
indicate that nanodust impact associated clouds strongly affect the
photoelectron environment of the antenna. In the scenario proposed by Zaslavsky
et al the photoelectrons emitted by the sunlight exposed surface of the antenna
are temporarily hindered to fall back onto it because of the presence of
the cloud's perturbing field. The resulting net photoelectron current is strong
enough to allow for a fast positive charging of the antenna which is compatible 
with the measured field intensities. The bottom line is that the presented
model is not directly applicable to the case of plasma clouds 
generated by nanodust impacts on spacecraft as neither the spacecraft
potential nor the surrounding photoelectrons have been considered. The
model is however expected to be applicable in case of nanodust impacts on
uncharged targets.

\section*{References}


\end{document}